# Deploying and Evaluating Multiple Deep Learning Models on Edge Devices for Diabetic Retinopathy Detection

Akwasi Asare [1], Dennis Agyemanh Nana Gookyi [2,*], Derrick Boateng [3], Fortunatus Aabangbio Wulnye [4]

[1] *Department of Computer Science, Ghana Communication Technology University, Accra, Ghana; nasare34@yahoo.com*

[2] *Electronics Division, Institute for Scientific and Technological Information, Council for Scientific and Industrial Research, Accra, Ghana; dennisgookyi@gmail.com*

[3] *College of Health Science and Environmental Engineering, Shenzhen Technology University, Shenzhen, 518188, China; derrickboateng@sztu.edu.cn*

[4] *Department of Telecommunication Engineering, Kwame Nkrumah University of Science and Technology, Kumasi, Ghana; fortunatuswulnye@outlook.com*

Corresponding Author: Dennis Agyemanh Nana Gookyi[2]

**Abstract:** Diabetic Retinopathy (DR), a leading cause of vision impairment in individuals with diabetes, affects approximately 34.6% of diabetes patients globally, with the number of cases projected to reach 242 million by 2045. Traditional DR diagnosis relies on the manual examination of retinal fundus images, which is both time-consuming and resource intensive. This study presents a novel solution using Edge Impulse to deploy multiple deep learning models for real-time DR detection on edge devices. A robust dataset of over 3,662 retinal fundus images, sourced from the Kaggle EyePACS dataset, was curated, and enhanced through preprocessing techniques, including augmentation and normalization. Using TensorFlow, various Convolutional Neural Networks (CNNs), such as MobileNet, ShuffleNet, SqueezeNet, and a custom Deep Neural Network (DNN), were designed, trained, and optimized for edge deployment. The models were converted to TensorFlowLite and quantized to 8-bit integers to reduce their size and enhance inference speed, with minimal trade-offs in accuracy. Performance evaluations across different edge hardware platforms, including smartphones and microcontrollers, highlighted key metrics such as inference speed, accuracy, precision, and resource utilization. MobileNet achieved an accuracy of 96.45%, while SqueezeNet demonstrated strong real-time performance with a small model size of 176 KB and latency of just 17 ms on GPU. ShuffleNet and the custom DNN achieved moderate accuracy but excelled in resource efficiency, making them suitable for lower-end devices. This integration of edge AI technology into healthcare presents a scalable, cost-effective solution for early DR detection, providing timely and accurate diagnosis, especially in resource-constrained and remote healthcare settings.

**Keywords**: Diabetic Retinopathy, Edge Impulse, Deep Learning, Microcontroller Units, TensorFlow, Model Quantization, Edge AI

# 1. INTRODUCTION

Diabetes is a significant global health challenge, with rates rising worldwide over the past two decades. One of the major complications of diabetes is Diabetic Retinopathy (DR), a severe eye condition that can lead to vision loss in adults (Maqsood and Gupta, 2022). DR is caused by damage to the blood vessels in the retina, leading to swelling and leakage, which can impair vision (Saeed, Hussain and Aboalsamh, 2021). Approximately 34.6% of individuals with diabetes develop DR, making it the leading cause of vision loss among working-age adults (Li *et al.*, 2023). By 2045, it is projected that there will be 242 million cases of DR and 71 million instances of advanced DR (Li *et al.*, 2023). Given the progressive nature of DR, which ranges from mild to severe stages, early detection and treatment are crucial for preventing vision loss and can reduce blindness by up to 95% (Burton et al., 2021; Grauslund, 2022). Traditionally, ophthalmologists diagnose DR by manually examining digital retinal fundus images and performing laboratory testing, processes that are both time-consuming and resource intensive. In response to this challenge, deep learning has emerged as a powerful tool for the automated detection and classification of DR, offering significant time and cost savings while improving the efficiency of hospital and clinic operations (Liang, Li and Deng, 2019). Deep learning uses artificial neural networks to extract high-level features from input data. It has been successfully applied in various domains, including image recognition, speech processing, and natural language understanding, and is increasingly being used to assist medical professionals in analyzing medical images (Liang, Li and Deng, 2019). Deep learning can revolutionize healthcare by improving diagnostic accuracy, accelerating drug discovery, and enabling personalized treatment (Kapoor and Arora, 2022; Onyema *et al.*, 2022; Emami and Javanmardi, 2023). However, in medical settings, deploying deep learning models for tasks like diabetic retinopathy detection is often hindered by the limited computational power of edge devices used in clinics or remote settings. Edge AI, which processes data directly on these local devices rather than relying on cloud services, provides a promising solution. This approach reduces latency and bandwidth requirements, enhances data privacy, and enables real-time decision-making, crucial for timely and effective disease diagnosis.

A key focus of our study is examining the resource utilization of various models and assessing their suitability for low-cost hardware devices like microcontrollers. This study places a significant emphasis on optimizing model performance for edge computing environments, ensuring that accurate diabetic retinopathy detection can be achieved without the need for expensive hardware. Given the financial constraints often faced in healthcare settings, particularly in resource-limited areas, the evaluation of model performance on low-cost edge devices will provide insights into the feasibility of implementing real-time diabetic retinopathy screening solutions, paving the way for improved healthcare delivery in underserved populations.

This paper tackles these challenges by leveraging Edge Impulse to deploy multiple deep learning models on a single edge device, offering a solution that seamlessly integrates into clinical or remote healthcare settings. Our study begins with the compilation of an extensive dataset containing thousands of retinal fundus images, categorized into various stages of diabetic retinopathy. This dataset is meticulously curated and enhanced using preprocessing techniques such as data splitting for training and testing, resizing, normalization, and augmentation (including rotation, flipping, and contrast adjustments) to create a robust and comprehensive resource for model training. Next, our approach focuses on the design and training of convolutional neural network (CNN) architectures within the TensorFlow framework, aiming to achieve high accuracy in diabetic retinopathy detection while optimizing the models for efficient deployment on edge devices. A range of CNN architectures, including MobileNet, ShuffleNet, SqueezeNet, and a custom Deep Neural Network (DNN), are explored and fine-tuned to strike a balance between predictive accuracy and computational efficiency. To ensure these models are compatible with the limited resources of edge devices, they are converted to the TensorFlowLite format and undergo 8-bit integer quantization. This process significantly reduces model size and enhances inference speed, which is essential for deploying models on edge hardware with restricted processing power and memory. During the deployment phase, we generate the necessary files and configure the Edge Impulse platform to support the simultaneous operation of multiple models on a single device. This approach not only broadens the capabilities for detecting various stages of diabetic retinopathy but also demonstrates the practical application of edge AI in real-world healthcare environments. Performance tests on edge hardware platforms, such as smartphones and microcontroller units, offer critical insights into the models' real-world performance (Garcia-Perez *et al.*, 2023). Important metrics, such as processing speed, accuracy, and resource consumption, are measured to determine the models' reliability and efficiency (Gookyi *et al.*, 2024). The diabetic retinopathy stages assessed include Moderate Diabetic Retinopathy, Proliferative Diabetic Retinopathy, Severe Diabetic Retinopathy, Mild Diabetic Retinopathy, and No Diabetic Retinopathy. **Figure 1** illustrates the various stages of DR.

The main contributions of this study are as follows:

- Developed and organized a detailed dataset of retinal fundus images representing various stages of diabetic retinopathy, serving as a solid foundation for model training.
- Designed and trained several convolutional neural network (CNN) models using TensorFlow to classify the different stages of diabetic retinopathy.
- Enhanced CNN models for edge deployment by converting them to TensorFlowLite and applying INT8 quantization, making them more suitable for devices with limited computational resources.
- Successfully implemented multiple deep learning models on a single edge device through the Edge Impulse platform, demonstrating the potential for real-time diabetic retinopathy detection directly on the device.
- Conducted a comprehensive analysis of the resource utilization of various deep learning models, assessing their compatibility with low-cost hardware devices like microcontrollers, and performed thorough performance assessments on edge hardware, evaluating key factors like inference speed, accuracy, and resource usage to confirm the models' effectiveness in real-world healthcare applications.

The subsequent sections of this study are structured as follows: Section 2 presents a detailed literature review, highlighting existing methods and approaches for diabetic retinopathy detection, including deep learning models and their deployment on edge devices. Section 3 describes the methodology used in this study, which involves the preparation of a retinal fundus image dataset, designing and training various convolutional neural network (CNN) architectures, and converting models into TensorFlowLite for edge deployment. Section 4 outlines the deployment process using Edge Impulse, covering model quantization, the generation of deployment files, and the configuration of edge devices for real-time diabetic retinopathy detection. In Section 5, the performance of the deployed models is evaluated across different edge hardware platforms, focusing on inference speed, accuracy, and resource consumption. Finally, Section 6 concludes the paper by summarizing the key findings and discussing the implications of edge AI in healthcare, specifically in improving access to diabetic retinopathy screening in resource-limited settings.

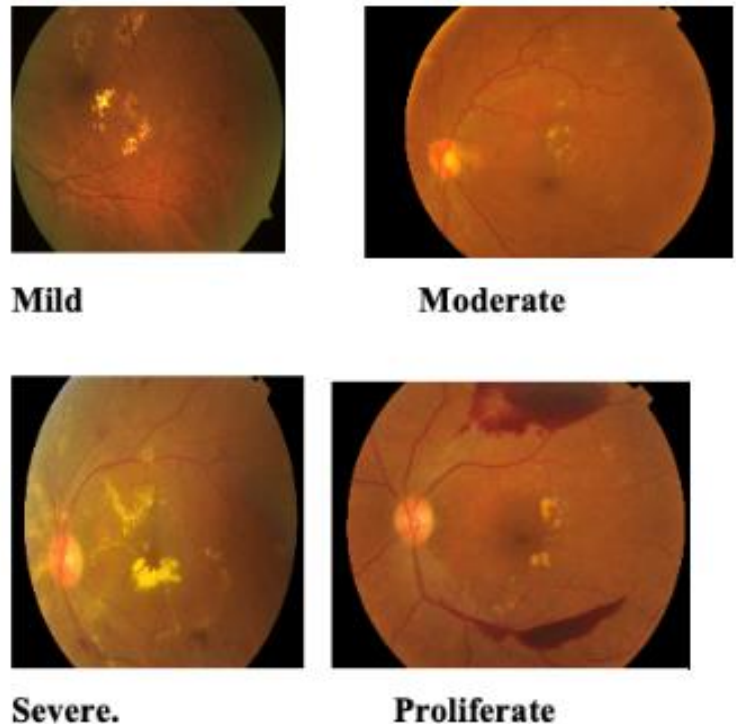

Figure 1. Different stages of DR (Emma *et al.*,2015).

## 2. LITERATURE REVIEW

Diabetic retinopathy poses a significant threat to global health, particularly affecting individuals with diabetes, and is a leading cause of vision loss in adults (Liang, Li and Deng, 2019). Early detection and accurate diagnosis are critical for effective management of the disease, helping to prevent severe visual impairment and blindness (Li *et al.*, 2023). Recent advancements in computer vision and deep learning have paved the way for automated systems capable of detecting diabetic retinopathy from retinal fundus images with high accuracy. Convolutional neural networks (CNNs) have emerged as a powerful tool in this domain, enabling precise classification of the different stages of diabetic retinopathy based on visual data. With the increasing need for real-time, on-device diagnostics, edge computing has become an integral component of these systems. Platforms such as Edge Impulse facilitate the deployment of multiple deep learning models on a single edge device, significantly enhancing the ability to perform real-time diagnostic analysis in clinical or remote settings. This paper reviews the current methodologies for diabetic retinopathy detection and highlights the limitations of existing approaches (Gookyi *et al.*, 2024).

Recent studies have explored various CNN architectures and their deployment on edge devices to improve detection accuracy and operational efficiency. For instance, Lekshmi and Rajathi (2019), implemented a system for the detection of glaucoma and diabetic retinopathy using artificial neural networks (ANNs) and image processing techniques on a Raspberry Pi (Lekshmi and Rajathi, 2019). However, their work focuses solely on Raspberry Pi-based processing, without investigating the system's adaptability to other hardware platforms, limiting its scalability for larger clinical deployments. Moreover, the use of only 90 images, which is a small dataset further limits the generalizability of the results and may not represent the variability of real-world image quality and conditions. In contrast, our study explores multiple CNN architectures to improve model generalization and assesses performance across different edge devices for broader applicability. We also incorporate thorough testing and validation to reduce overfitting and ensure strong performance on unseen data. This detailed evaluation allows us to identify the most effective and efficient architectures for practical deployment.

Deshmukh *et al*. (2023) developed a remote DR screening system utilizing IoT for data collection and machine learning algorithms for real-time image analysis on edge devices (Deshmukh *et al.*, 2023). However, while their system demonstrates improved accessibility and operational efficiency, it primarily focuses on edge device deployment without a detailed comparison of performance across other platforms, potentially limiting its scalability in diverse healthcare settings. Furthermore, the study emphasizes accessibility but lacks a deep exploration of machine learning model optimization for varied imaging conditions, which may affect generalizability. In contrast, our research evaluates multiple machine learning models and thoroughly tests their performance across various edge devices, ensuring robust model generalization. Our comprehensive testing on unseen data reduces overfitting, making our system more suitable for broader clinical deployment.

Vidhya Lavanya et al. (2020)also conducted a study on the detection and Classification of Diabetic Retinopathy using Raspberry Pi with ImageNet CNN, achieving an overall weighted accuracy of 67.37% (Vidhya Lavanya *et al.*, 2020). However, their study demonstrated a lower accuracy and lacked a comprehensive evaluation of key performance metrics such as sensitivity, specificity, F1 score, and AUC. Additionally, it did not provide an in-depth assessment of real-time performance, including latency, inference speed, and resource utilization which are critical for clinical diabetic retinopathy screening. The reliance on Raspberry Pi, a low-cost but resource-constrained device, limits the complexity of models that can be deployed, affecting both detection accuracy and speed. Our study addresses this by conducting a comprehensive evaluation of key performance metrics such as sensitivity, specificity, F1 score, and AUC and exploring the real-time performance metrics on MCUs and GPUs such as latency, inference speed, and resource utilization.

Singh *et al*. (2023) developed a system for detecting diabetic retinopathy using the YOLOv8 deep learning model integrated with the Jetson Nano Developer Kit (Singh *et al.*, 2023) While their system demonstrates high accuracy (93.77% on training data), the use of a relatively small dataset (IDRID, consisting of only 122 images) limits the scalability and generalizability of their findings. Additionally, the study focuses on the Jetson Nano, which may restrict its adaptability to other hardware platforms, further hindering broader clinical application. Furthermore, the paper does not address real-time performance or testing under varied network conditions, which are critical for evaluating inference speed, latency, and system responsiveness in real-world diabetic retinopathy detection scenarios. On the other hand, our research not only explores multiple CNN architectures across different edge devices but also rigorously evaluates model performance under diverse network conditions and user scenarios. This ensures optimal real-time inference speed and system responsiveness, providing a more comprehensive solution for scalable, real-world deployments of diabetic retinopathy screening systems.

Patel *et al*. (2021) developed a browser-based application called DiaRet for the grading of Diabetic Retinopathy with integrated gradients (Patel *et al*., 2021). While browser-based solutions provide accessibility, their system faces limitations in terms of reporting on the model's accuracy, dataset source, scalability, latency, and computational performance on edge devices. In contrast, our research emphasizes the model's performance reporting and edge deployment, which allows for faster inference, reduced dependency on internet connections, and greater flexibility in resource-constrained environments such as rural or underdeveloped areas.

Nurmalasari et al. (2022) evaluated six CNN architectures—AlexNet, DenseNet121, InceptionV3, ResNet50, VGG16, and Xception for diabetic retinopathy (DR) severity classification. VGG16 achieved the highest accuracy of 77% (Nurmalasari *et al.*, 2022). The models were assessed with and without optimization techniques such as ADAM, SGD, and RMSprop, underscoring the importance of model optimization in DR classification. While their study provided a comprehensive comparative performance analysis, it did not address the trade-offs between model complexity and real-time applicability. Larger CNN architectures can be resource-intensive, making them less suitable and difficult for edge deployment. Our research aims to optimize both model size and inference speed by quantizing models to INT8. We have implemented techniques to enhance both model performance and inference efficiency, ensuring that our models are suitable for edge deployment without sacrificing accuracy. Additionally, our approach emphasizes rigorous testing and validation, ensuring that the models are reliable and effective for real-world diabetic retinopathy detection, bridging the gap between high accuracy in controlled settings and practical application.

Satyananda *et al.* (2019) developed a diagnostic system utilizing multiple machines learning methods, including Probabilistic Neural Networks (PNN), Support Vector Machines (SVM), Bayesian Classification, and K-Means Clustering, to analyze retinal fundus images (Satyananda *et al.,* 2019). However, their study was constrained by a limited dataset of 300 images, which may not sufficiently reflect the diversity and complexity of real-world retinal image data. Additionally, the study's focus was on comparing algorithm performance without exploring how these models could be optimized for a range of hardware platforms, limiting its potential for real-time, large-scale clinical applications. However, our research focuses on optimizing and adapting machine learning models for deployment on various embedded systems, improving both scalability and generalizability. By utilizing a more extensive dataset of 3,662 images and testing across different hardware environments, such as microcontroller units (MCUs) and GPUs, we ensure that our models are robust and applicable in real-world clinical settings. This thorough testing also addresses overfitting concerns, resulting in models that maintain high performance when applied to unseen data.

Matthew *et al.* (2023) developed a DR diagnosis system using the EfficientNet-B0 architecture, implemented through transfer learning, for real-time classification of retinal images on Android devices (Matthew, Gunawan and Kurniadi, 2023). Their model achieved a classification accuracy of 91.85% across three classes: No DR, Non-Proliferative DR, and Proliferative DR. Despite its promising performance, the system's focus on mobile deployment and a dataset with imbalanced classes presents challenges for generalizability and scalability in larger clinical settings. Additionally, the study primarily explores a single mobile-based platform without assessing adaptability to other hardware environments. In contrast, our research evaluates multiple CNN architectures and tests their performance on various edge devices to ensure broader applicability and improved generalization. Through comprehensive validation, we aim to reduce overfitting and enhance model performance for practical deployment in diverse clinical environments.

Kumar et al. (2023) utilized various architectures for mobile-based classification and detection of diabetic retinopathy using TinyML, achieving accuracies of 79.8% for ResNet-18, 78.6% for ResNet-34, 81.47% for ResNet-50, 82.37% for ResNet-101, 77.38% for RegNet, and 78.74% for WideResNet. While the study employed techniques like quantization, pruning, and model distillation to reduce model complexity, it did not conduct rigorous testing across various mobile devices with different hardware capabilities. Our research fills this gap by extensively testing models on various edge devices, from high-end smartphones with AI chips to low-end models, ensuring broad applicability. Furthermore, Kumar et al. (2023) did not provide a detailed analysis of real-time performance, scalability, or model behavior in different mobile environments. In contrast, our study emphasizes real-time testing, evaluating model performance under various network conditions and user scenarios to ensure optimal inference speed, latency, and system responsiveness for real-world diabetic retinopathy detection.

The "MAILOR AI study" Rogers *et al.* (2019) developed the Pegasus AI system for diabetic retinopathy (DR) detection using handheld fundus cameras (Rogers *et al.*, 2019). The system achieved an AUROC of 94.3% for Proliferative DR (PDR) but a lower AUROC of 89.4% for Referable DR (RDR), compared to 98.5% with a desktop camera. This decline in accuracy was attributed to lower image quality and inconsistencies between the grading systems used (Scottish DR grading vs. ICDR scale). However, our study improves model generalization by testing multiple AI architectures across different edge devices and employing robust standardized machine learning protocols like cross-validation and regularization to avoid overfitting. Additionally, we incorporate image preprocessing techniques to enhance data quality, leading to improved accuracy across devices. This comprehensive evaluation ensures broader applicability and scalability for real-world clinical deployments.

This thorough review of the literature highlights key advancements and challenges in diabetic retinopathy detection using deep learning models. While progress has been made, there is still a need for more efficient CNN architectures, diverse datasets, and better deployment strategies on edge devices. Our research addresses these gaps by evaluating model performance under various network conditions and user scenarios to ensure optimal inference speed, latency, and system responsiveness for real-world diabetic retinopathy detection while also advancing current methodologies to create more robust and practical solutions.

## 3. METHOD AND MATERIALS

The proposed methodology follows a deep learning-based diabetic retinopathy (DR) detection pipeline optimized for deployment on edge devices. Retinal fundus images serve as inputs to lightweight convolutional neural networks (CNNs) (MobileNet, ShuffleNet, SqueezeNet), which are optimized using TensorFlow Lite quantization. The models run efficiently on microcontrollers and edge hardware, classifying DR severity into Normal, Mild, Moderate, Severe, or Proliferative DR, enabling real-time, AI-driven DR screening in resource-limited settings. The conceptual framework of our models are illustrated in **Figure 2a**

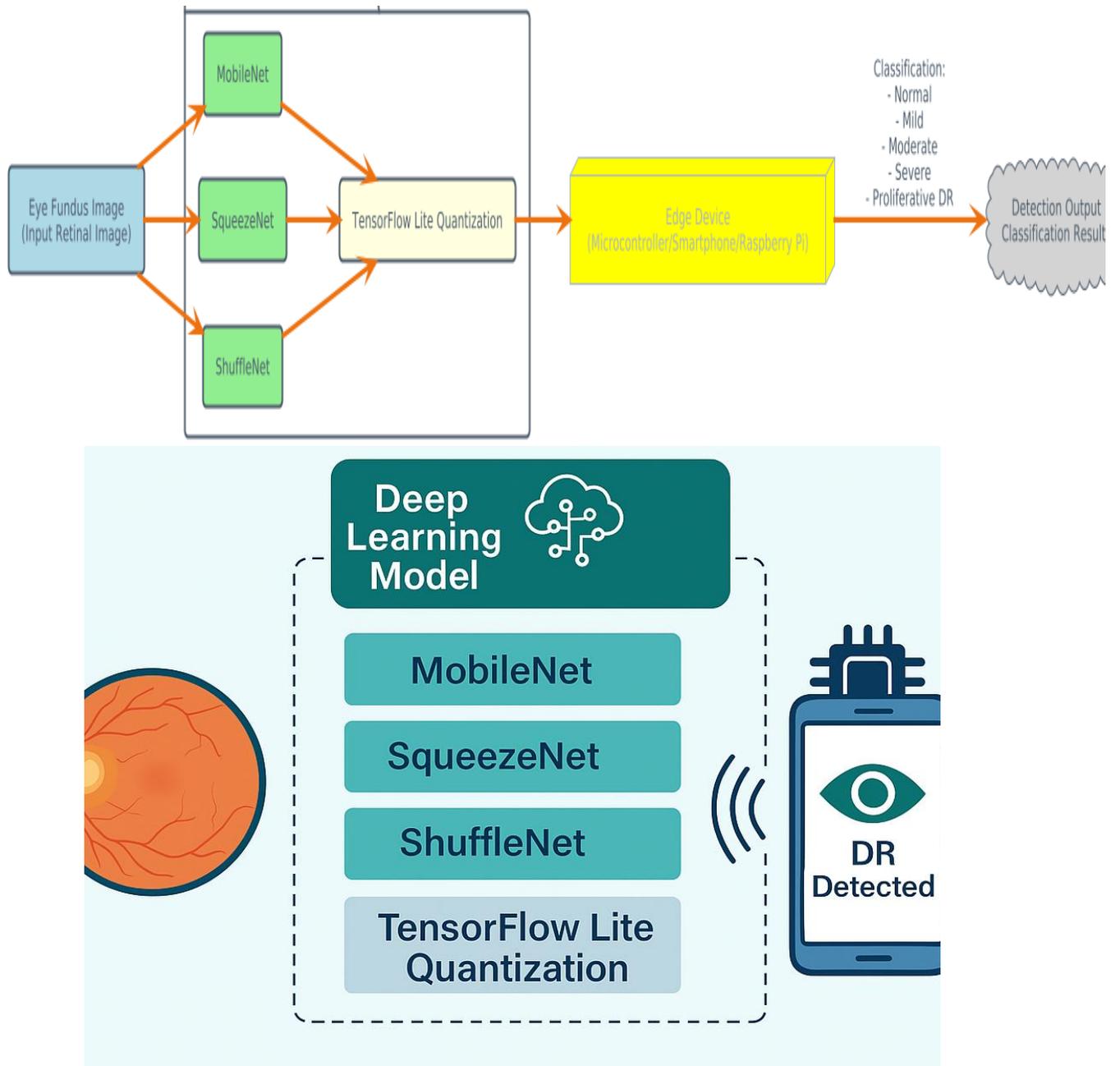

**Figure 2a**. Conceptual Framework of our proposed models

To tackle the challenges of real-time diabetic retinopathy detection in clinical and remote healthcare environments, we developed a structured methodology that integrates deep learning techniques with edge computing. Our solution employs TensorFlow for training models and Edge Impulse for efficient deployment on edge devices, ensuring the models are both accurate and capable of real-time inference on devices with limited resources. The methodology is divided into two key phases: the TensorFlow phase, which includes dataset preparation, model design, training, and conversion to TensorFlowLite; and the Edge Impulse phase, which involves deploying the trained and quantized models to edge devices and assessing their performance in practical healthcare scenarios. The methodology phases are illustrated in **Figure 2**.

In the TensorFlow phase, we begin by assembling a comprehensive dataset of retinal fundus images and categorizing them into the different stages of diabetic retinopathy, which is essential for training effective deep learning models. To enhance the model's ability to generalize to new, unseen data, data augmentation techniques such as rotation, flipping, and contrast adjustments are applied. Various convolutional neural network (CNN) architectures are then designed and trained for the classification of different stages of diabetic retinopathy, with hyperparameters fine-tuned to achieve optimal performance. Figure 3 illustrates the TensorFlow phase, covering dataset preparation, model design, training, and conversion, and the Edge Impulse phase, focusing on deploying the models to edge devices and evaluating their real-world performance. Following training, the models are converted into TensorFlowLite (TFLite) format and quantized to 8-bit integers. This process is crucial for minimizing model size and computational requirements, making them suitable for deployment on edge devices with limited processing capabilities.

In the Edge Impulse phase, deployment files are created, deployment platforms are configured, and the models' performance is thoroughly evaluated across various edge devices. This ensures the models retain both high accuracy and efficiency when deployed in real-world healthcare scenarios. By carefully managing each stage of model development and deployment, our approach delivers a practical and scalable solution for diabetic retinopathy detection, facilitating timely and precise diagnosis, which is critical for preventing vision loss. The detailed steps of this methodology are illustrated in **Figure 3**.

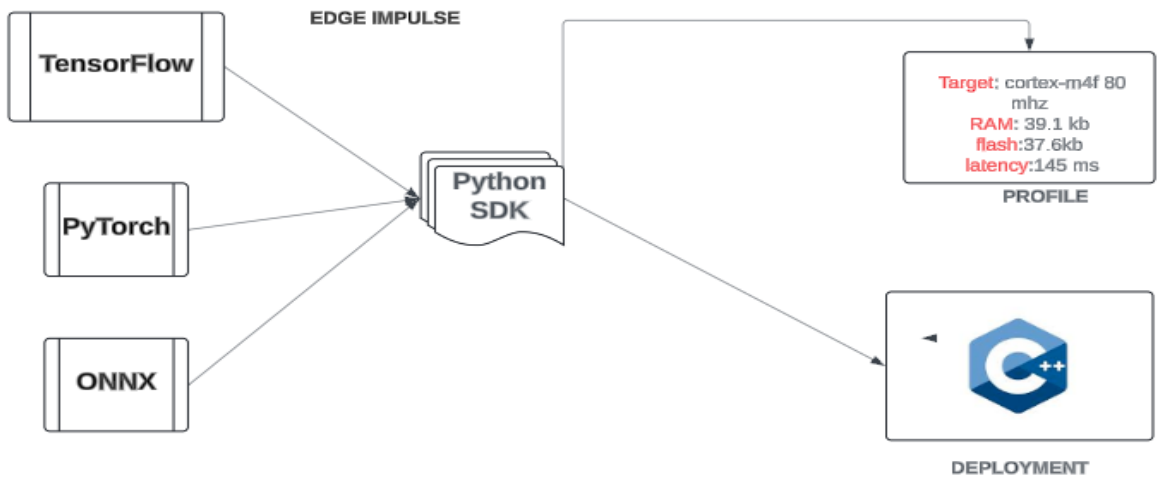

Figure 2. TensorFlow for model training and Edge Impulse for deployment and evaluation.

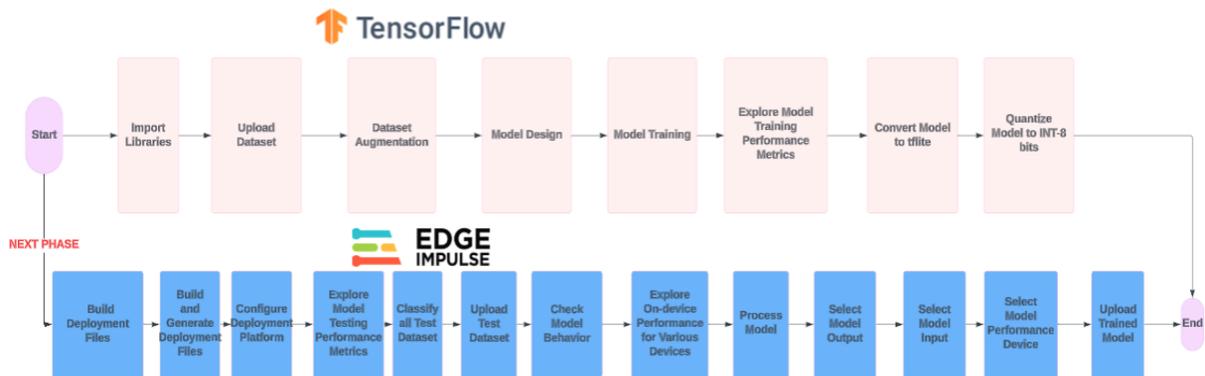

Figure 3. The phases of the proposed method.

**3.1. TensorFlow**

3.1.1. Dataset Acquisition and Augmentation

The dataset used for training the deep learning models in this study was a collection of retinal fundus images provided by the EyePACS organization, hosted on Kaggle (Emma *et al*., 2015). This dataset offers a comprehensive range of retinal images, capturing different stages of diabetic retinopathy (DR). For our analysis, we selected a subset of 3,662 images from the total 35,126 images in the dataset and classified them into five categories based on DR severity: No DR, Mild, Moderate, Severe, and Proliferative DR. Specifically, the dataset comprised 1,805 images with No DR, 370 images with Mild DR, 999 images with Moderate DR, 193 images with Severe DR, and 295 images with Proliferative DR. Sample images from this dataset are presented in **Figure 4**.

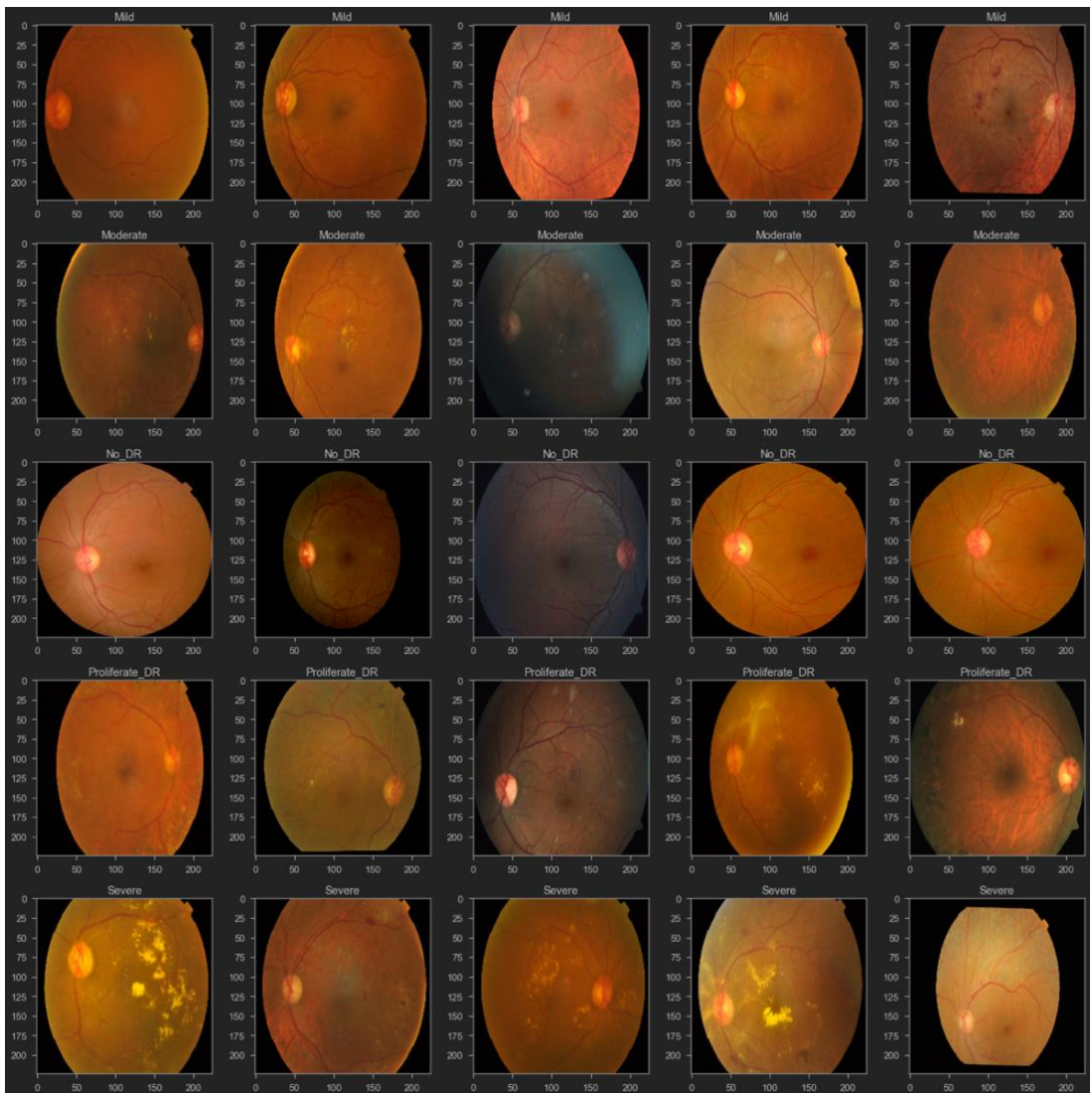

Figure 4. Samples of the dataset images.

This selection provided a balanced distribution across the different severity levels, which was crucial for training reliable deep-learning models. We further applied data augmentation techniques such as rotation, flipping, and contrast adjustments to enhance the model's ability to generalize to new, unseen data. The dataset was divided into training and validation sets, with an 80/20 split, with the training set used for model development and the validation set for performance evaluation and fine-tuning of hyperparameters. The distribution of the dataset is shown in **Figure 5**.

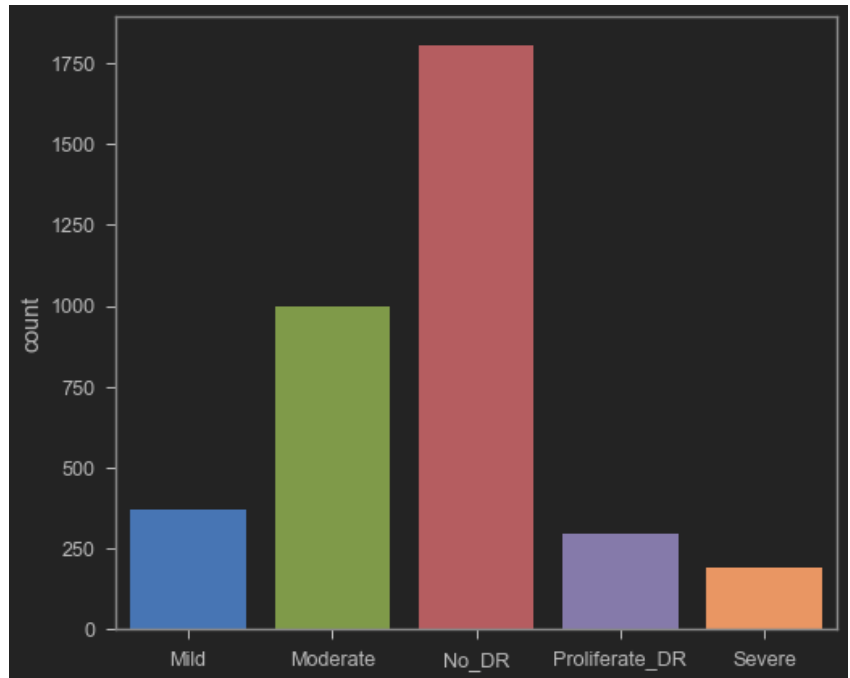

Figure 5. Distribution of Diabetic Retinopathy disease dataset.

3.1.2. Models Design and Training

During the model design phase, we evaluated several state-of-the-art convolutional neural network (CNN) architectures, chosen for their balance between accuracy, computational efficiency, and suitability for deployment on edge devices with limited resources. The architectures considered included MobileNet, ShuffleNet, SqueezeNet, and a custom deep neural network (DNN) specifically designed for the task of diabetic retinopathy detection.

- MobileNet: We employed MobileNet, a compact and streamlined neural network architecture designed for mobile and embedded systems. It utilizes depthwise separable convolutions to minimize computational requirements (Prasetyo *et al.*, 2023). The architecture, illustrated in Figure 6, consists of depthwise and pointwise convolution layers, as well as batch normalization, ReLU activations, global average pooling, and dropout regularization. Thanks to its computational efficiency, MobileNet is well-suited for applications with limited resources, such as real-time diabetic retinopathy detection on edge devices(Suriyal, Druzgalski and Gautam, 2018).

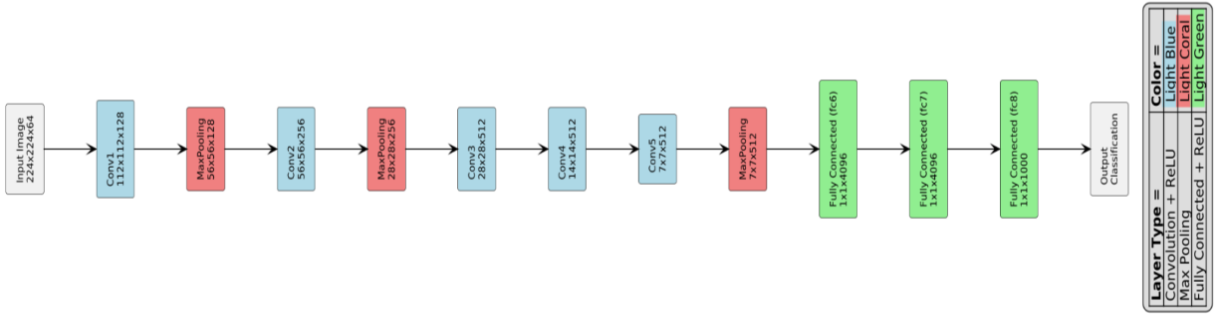

Figure 6. MobileNet architecture.

- ShuffleNet: ShuffleNet is also a compact and lightweight convolutional neural network designed for mobile platforms, utilizing pointwise group convolutions and channel shuffling to reduce computational complexity while preserving accuracy (Zhang *et al.*, 2017). The channel shuffle mechanism ensures effective feature learning across different groups, making ShuffleNet particularly well-suited for resource-limited environments without compromising performance in tasks such as image classification. The architecture of the ShuffleNet model is illustrated in **Figure 7**.

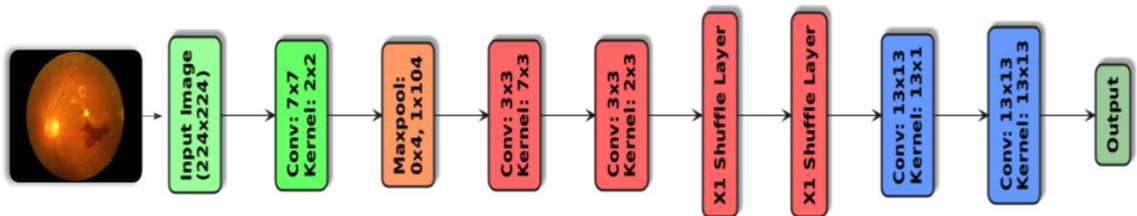

Figure 7. ShuffleNet architecture.

- SqueezeNet: SqueezeNet is an efficient neural network architecture designed to achieve accuracy comparable to AlexNet while utilizing fewer parameters. It employs "squeeze" and "expand" layers to minimize the parameter count, making it highly optimized for environments with constrained memory and computational power (Iandola *et al.*, 2016; Nanthini *et al.*, 2023) This design allows SqueezeNet to perform exceptionally well on image classification tasks while remaining suitable for deployment on resource-limited devices. In this study, we further enhanced SqueezeNet's deployment efficiency by optimizing its performance specifically for edge devices, ensuring both speed and accuracy in real-time diabetic retinopathy detection. **Figure 8** depicts the architecture of SqueezeNet.

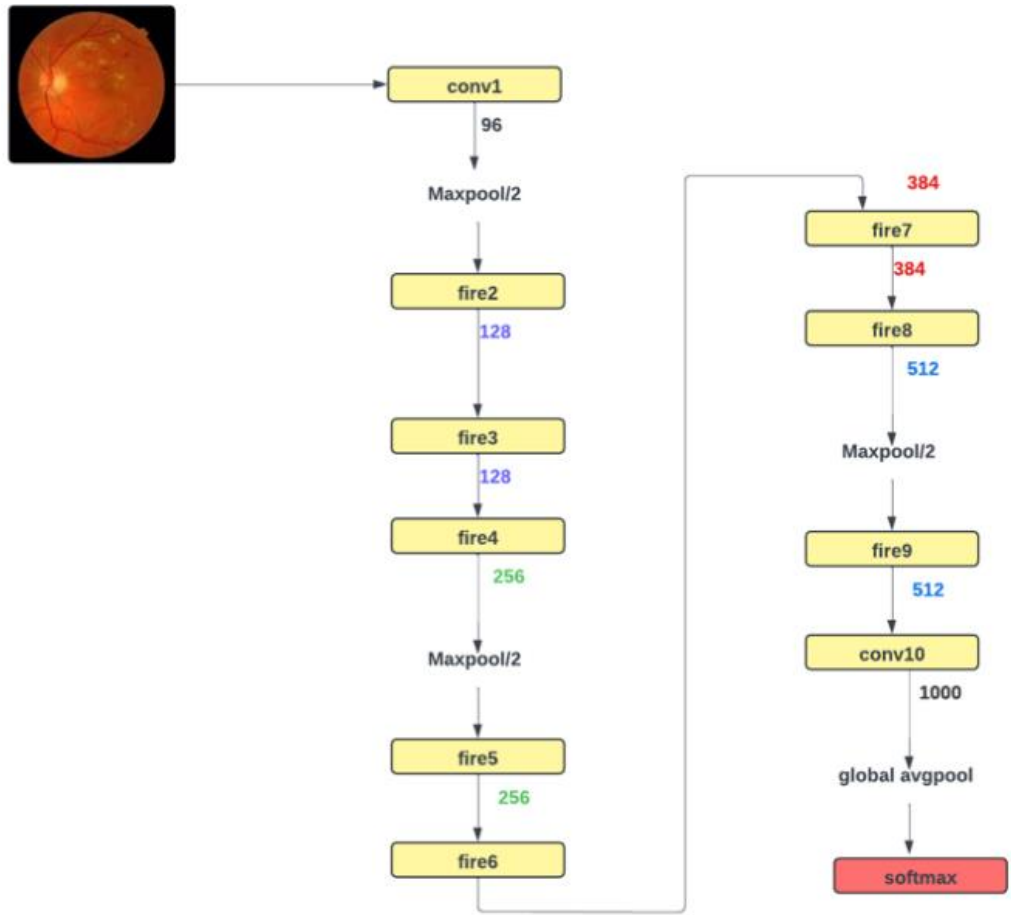

Figure 8. SqueezeNet architecture.

- Custom DNN: The custom deep neural network (DNN) architecture consists of multiple convolutional (Conv2D) layers paired with max-pooling (MaxPooling2D) layers, followed by a flattening layer that bridges the transition from convolutional layers to fully connected layers. The convolutional layers use different filter sizes and quantities to capture hierarchical features from the input data, while max-pooling is employed to down-sample spatial dimensions, preserving key features. The flattening layer converts the output into a vector, preparing it for input into the fully connected layers for classification. This architecture is tailored for classification tasks and, in this study, has been adapted to classify various stages of diabetic retinopathy from retinal fundus images. By fine-tuning the architecture, we achieved high accuracy in detecting and categorizing the severity of the disease. **Figure 9** illustrates the structure of the custom DNN model.

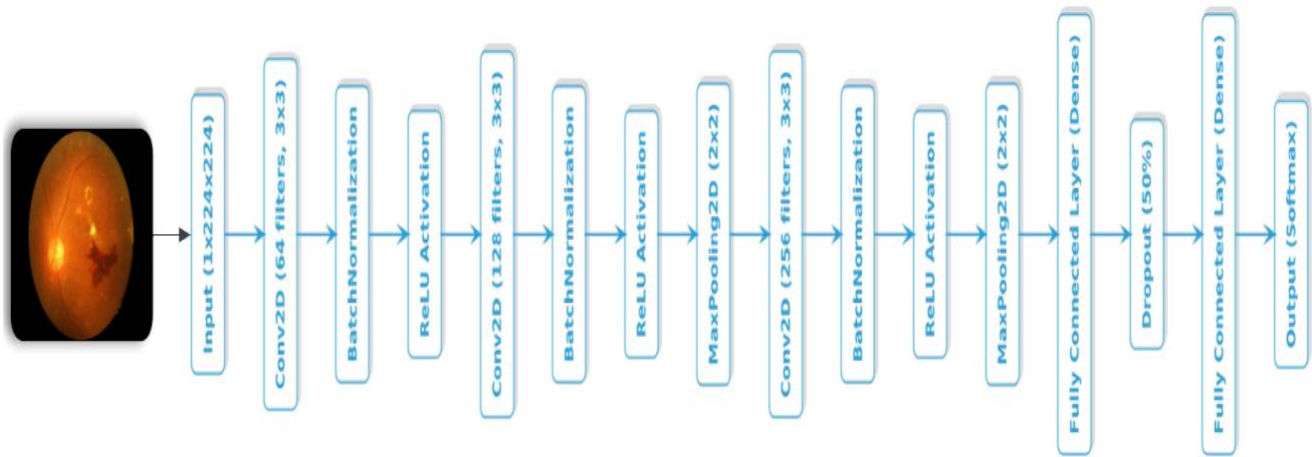

Figure 9. The architecture of Custom DNN.

3.1.3. Model Training and Performance Evaluation

Each model was trained using the prepared dataset, with hyperparameters carefully tuned to maximize performance. The training process incorporated several techniques, including learning rate annealing and batch normalization, to prevent overfitting and ensure convergence. All the original image sizes were maintained at 224×224 pixels to preserve and standardize the input size. The training utilized a learning rate of 0.001 with the Adam optimizer, beta1 (0.9) and beta2 (0.999), a batch size of 32, and a total of 20 epochs. Beta1 (0.9) represents the exponential decay rate for the first-moment estimate, which is essentially the moving average of the gradients. Setting beta1 to 0.9 means that the optimizer places 90% weight on the previous gradient value and 10% on the current gradient. This helps smooth out fluctuations in the gradient updates, preventing sudden changes that could destabilize training. Beta2 (0.999) controls the exponential decay rate for the second moment estimate, which captures the moving average of the squared gradients. A high value of 0.999 means that the optimizer gives more weight to past squared gradients, allowing it to adjust the learning rate more effectively over time. This helps the optimizer handle the varying scale of gradients during training.

Post-training, each model's performance was assessed using common evaluation metrics such as accuracy, precision, recall, and F1-score, calculated on a reserved validation set. This allowed us to determine the most effective models for conversion and deployment. The models were then converted into TensorFlowLite (TFLite) format to ensure compatibility with edge devices. During conversion, INT8 quantization was applied, significantly reducing the model size and computational requirements. While this quantization process slightly reduced precision, it was critical for maintaining high inference speeds and low power consumption, making the models ideal for real-time operation on resource-limited edge devices.

**3.2 Edge Impulse**

Edge Impulse is a versatile platform designed to simplify the deployment of machine learning models on edge devices, making it highly suitable for medical applications such as diabetic retinopathy detection. It facilitates efficient model conversion and optimization while supporting real-time analysis in resource-constrained environments, such as point-of-care devices or mobile health systems. One of its key strengths is streamlining the entire workflow from data collection and model training to deployment, while minimizing the computational load on edge devices. **Figure 10** depicts the complete cycle of training, optimizing, and deploying machine learning models using Edge Impulse, showcasing its potential for low-latency, on-device diabetic retinopathy screening. Additionally, its integration with popular machine learning frameworks and its user-friendly interface promote accessibility for both healthcare professionals and AI developers, advancing innovations in edge AI for medical diagnostics.

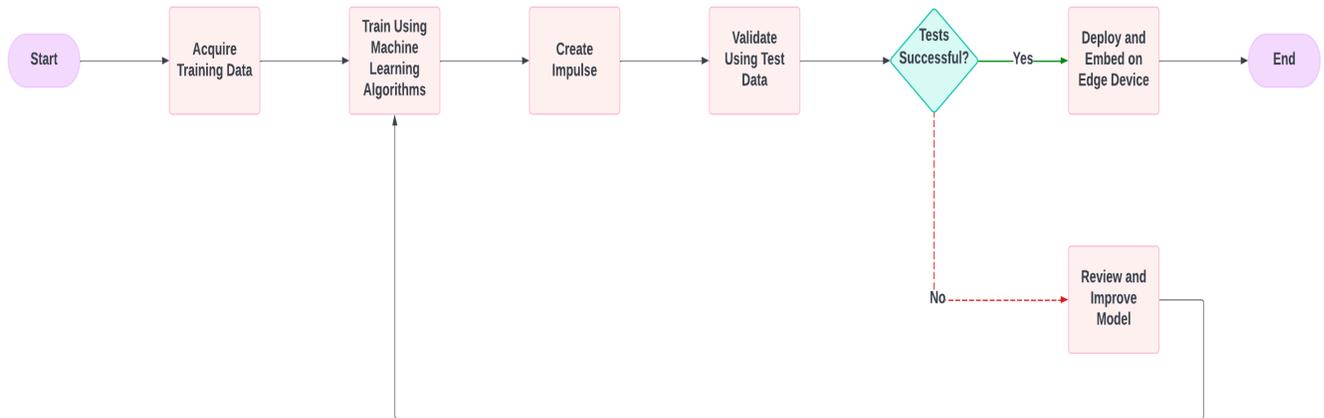

Figure 10. Edge Impulse architecture.

3.2.1 File Generation for Deployment
The next phase of our methodology focused on preparing the models for deployment using Edge Impulse, a platform designed to streamline the implementation of machine learning models on edge devices. After selecting the best-performing models from the TensorFlow segment, we utilized Edge Impulse to generate deployment files. This process involved converting the models into formats optimized for the specific hardware of our target edge devices, such as microcontrollers or single-board computers, commonly used in medical applications like diabetic retinopathy screening. Configuring the deployment platform required setting up the edge devices with the necessary software environments, including the installation of Edge Impulse's firmware and runtime. This setup ensured the edge devices could efficiently run TFLite models with all dependencies properly managed, allowing for real-time diabetic retinopathy detection in resource-limited environments.

3.2.2  Model Testing and Performance Assessment
Once the deployment files were created and the platform was set up, we conducted extensive tests on the target edge devices. This testing phase was critical for verifying the models' effectiveness in real-world conditions, where factors such as fluctuating lighting, device power limitations, and processing lags could significantly impact both inference accuracy and speed. We gathered performance metrics such as inference time, power usage, and on-device accuracy to determine the models' suitability for practical applications. Additionally, we analyzed memory usage and computational efficiency to ensure the models could operate continuously on edge devices without performance degradation. The entire test dataset was classified using the deployed models to evaluate their real-time performance in detecting diabetic retinopathy, accounting for varying degrees of disease severity. This thorough evaluation helped fine-tune the models for use in resource-limited clinical settings, ensuring they deliver both accuracy and reliability in real-world healthcare scenarios.

3.2.3. On-Device Performance Assessment

In the final phase of our methodology, we conducted a thorough assessment of the models' performance directly on the edge devices. This involved continuous monitoring to evaluate stability, responsiveness, and accuracy over extended periods. We carefully analyzed the models' behavior under various operational conditions, including fluctuating power levels and environmental variability, to ensure reliable performance in real-world healthcare settings, such as remote clinics for diabetic retinopathy screening. Any anomalies or performance issues identified during the on-device evaluation were addressed through iterative improvements in the model design or adjustments to the deployment configuration. The refined models were then re-uploaded to the Edge Impulse platform for further enhancement, ensuring that the final deployment was reliable, efficient, and suitable for real-time medical use. By following this rigorous approach, our goal was to provide a practical solution for diabetic retinopathy detection that is not only accurate but also deployable in resource-constrained environments, enabling healthcare providers to offer timely and effective screenings for diabetic patients.

## 4. RESULTS

In this section, we present the experimental results from evaluating various models designed for diabetic retinopathy detection. The experiments were carefully conducted using a variety of evaluation metrics, including accuracy, loss, and training time. The goal of this study was to assess the performance of each model and compare its effectiveness in accurately detecting diabetic retinopathy while distinguishing it from healthy retinal images. Through systematic analysis and comparison of the experimental outcomes, we sought to identify the strengths and limitations of each model, ultimately aiming to determine the most accurate and efficient model for deployment in real-world medical applications. The experimental results offer valuable insights into the performance of different models, which guide in selecting optimal solutions in the context of diabetic retinopathy screening.

4.1. **Model Results**

4.1.1. MobileNet

The MobileNet model demonstrated strong performance during training, achieving a high accuracy of 96.45% with a low training loss of 0.1096, as detailed in **Figure 11 (a)**. However, the validation accuracy dropped to 73.58%, with a corresponding validation loss of 1.2921. This gap between training and validation performance indicates potential overfitting, as the model struggles to generalize on the validation data. Despite its compact size of 3.4MB and efficient training time of 3,024.75 seconds, MobileNet's ability to generalize may benefit from further tuning, such as applying regularization techniques or augmenting the dataset to improve validation performance.

The evaluation of MobileNet across various hardware platforms revealed significant differences in latency, underscoring the importance of selecting appropriate devices for real-time diabetic retinopathy detection. On low-end MCUs, the model exhibited a substantial latency of 156,957 ms, making it impractical for time-sensitive applications. However, on high-end MCUs, the latency was reduced dramatically to 3,214 ms, offering a more feasible option for edge devices in constrained environments. AI accelerators further improved performance, with latency dropping to 536 ms, highlighting their potential for optimizing on-device inference. The most efficient results were achieved on traditional computing platforms, with a latency of 109 ms on CPUs and just 19 ms on GPUs, as shown in Table 1 and Table 2, demonstrating that more powerful hardware can significantly enhance real-time processing. These findings suggest that while low-end MCUs may struggle with the demands of MobileNet, higher-end MCUs, AI accelerators, and GPUs can provide viable solutions for real-time diabetic retinopathy screening, depending on the deployment environment and resource availability.

**Table 1: MobileNet MCU Performance Metrics**

| Device | Latency | EON Compiler RAM | EON Compiler ROM | TFLite RAM | TFLite ROM |
|---|---|---|---|---|---|
| Low-end MCU | 156,957 ms | 1.0M | 3.4M | 2.0M | 3.5M |
| High-end MCU | 3,214 ms | 1.0M | 3.4M | 2.0M | 3.5M |
| AI Accelerator | 536 ms | 1.0M | 3.4M | 2.0M | 3.5M |

**Table 2: MobileNet Microprocessor Performance**

| Device | Latency | Model Size |
|---|---|---|
| CPU | 109 ms | 3.5M |
| GPU or accelerator | 19 ms | 3.5M |

4.1.2. ShuffleNet

The ShuffleNet model showed moderate performance during training, achieving an accuracy of 72.96% with a training loss of 0.7548, as shown in **Figure 11 (b)**. The validation accuracy was slightly lower at 68.47%, with a corresponding validation loss of 0.8177. This discrepancy between training and validation results suggests that ShuffleNet might struggle to generalize effectively on unseen data. Despite its small model size of 68.5 KB and efficient training time of 425.06 seconds, ShuffleNet's performance could benefit from additional tuning or regularization to improve its generalization on validation data.

The evaluation of ShuffleNet across various hardware platforms revealed notable differences in latency, emphasizing the need to choose suitable devices for real-time diabetic retinopathy detection. On low-end MCUs, the model experienced a significant latency of 23,169 ms, making it unsuitable for time-sensitive applications. In contrast, high-end MCUs showed a much-improved latency of 476 ms, providing a more practical option for edge devices in constrained environments. AI accelerators further enhanced performance, reducing latency to 80 ms, demonstrating their potential for optimizing on-device inference. The most efficient results were seen with traditional computing platforms, achieving latencies of 15 ms on CPUs and just 3 ms on GPUs, as shown in Table 3 and Table 4, highlighting how more powerful hardware can greatly improve real-time processing capabilities. These findings indicate that while low-end MCUs may struggle with

ShuffleNet's demands, higher-end MCUs, AI accelerators, and GPUs can offer effective solutions for real-time diabetic retinopathy screening based on the deployment context and resource availability.

**Table 3: ShuffleNet MCU Performance Metrics**

| Device | Latency | EON Compiler RAM | EON Compiler ROM | TFLite RAM | TFLite ROM |
|---|---|---|---|---|---|
| Low-end MCU | 23,169 ms | 312.9K | 54.1K | 536.4K | 75.7K |
| High-end MCU | 476 ms | 312.9K | 68.5K | 536.5K | 92.8K |
| AI Accelerator | 80 ms | 312.9K | 68.5K | 536.5K | 92.8K |

**Table 4: ShuffleNet Microprocessor Performance**

| Device | Latency | Model Size |
|---|---|---|
| CPU | 15 ms | 33.2 K |
| GPU or accelerator | 3 ms | 33.2K |

4.1.3. SqueezeNet
The SqueezeNet model exhibited commendable performance during training, achieving a training accuracy of 72.37% with a training loss of 0.7543, as detailed in **Figure 11 (c)**. Validation results showed a slight decrease, with a validation accuracy of 72.59% and a validation loss of 0.7371. This close alignment between training and validation metrics suggests that the model is effectively generalizing on unseen data. With a model size of 176.1 KB and a training time of approximately 2,427.55 seconds, SqueezeNet demonstrates efficiency in both resource utilization and training duration. While the model performs well, further refinement could still be explored, such as incorporating regularization techniques or data augmentation strategies, to optimize performance further.

The evaluation of SqueezeNet across various hardware platforms also revealed significant differences in latency, highlighting the importance of selecting appropriate devices for real-time diabetic retinopathy detection. On low-end MCUs, the model exhibited a substantial latency of 241,592 ms, rendering it impractical for time-sensitive applications. In contrast, high-end MCUs showed a marked improvement with a latency of 4,946 ms, providing a more feasible option for edge devices in constrained environments. AI accelerators further enhanced performance, reducing latency to 825 ms, demonstrating their potential for optimizing on-device inference. The most efficient results were observed on traditional computing platforms, achieving latencies of 101 ms on CPUs and just 17 ms on GPUs, as shown in Table 5 and Table 6, illustrating how more powerful hardware can significantly enhance real-time processing capabilities. These findings suggest that while low-end MCUs may struggle with SqueezeNet's demands, higher-end MCUs, AI accelerators, and GPUs can offer effective solutions for real-time diabetic retinopathy screening, depending on the deployment context and resource availability.

Table 5: SqueezeNet MCU Performance Metrics

| Device | Latency | EON Compiler RAM | EON Compiler ROM | TFLite RAM | TFLite ROM |
|---|---|---|---|---|---|
| Low-end MCU | 241,592 ms | 3.1M | 168.1K | 4.6M | 197.0K |
| High-end MCU | 4,946 ms | 3.1M | 176.1K | 4.6M | 206.2K |
| AI Accelerator | 825 ms | 3.1M | 176.1K | 4.6M | 206.2K |

Table 6: SqueezeNet Microprocessor Performance

| Device | Latency | Model Size |
|---|---|---|
| CPU | 101 ms | 154.6 K |
| GPU or accelerator | 17 ms | 154.6K |

4.1.4. Custom DNN

The custom DNN model exhibited strong performance during training, achieving an impressive accuracy of 90.93% with a low training loss of 0.2454, as illustrated in **Figure 11 (d)**. However, the validation accuracy declined to 75.43%, accompanied by a validation loss of 1.1215. This disparity in performance between the training and validation sets suggests potential overfitting, where the model excels in fitting the training data but struggles to generalize effectively to the validation set. The notable divergence in accuracy and the elevated validation loss suggest that the model may be memorizing the training data rather than learning patterns that generalize well to new data. With a model size of 1.6MB and a training time of 2,239.12 seconds, this custom DNN demonstrates efficiency in terms of memory and computational resources, making it relatively lightweight for deployment. However, to enhance its generalization ability, several improvements can be considered. Utilizing regularization techniques such as dropout or L2 regularization could help mitigate overfitting by promoting the learning of more robust features. Furthermore, augmenting the dataset by introducing variability in the training data. For instance, through transformations like rotations, flips, or noise could aid the model in better generalizing on unseen data.

The evaluation of the custom DNN model across various hardware platforms revealed significant differences in latency, emphasizing the need for appropriate device selection for real-time diabetic retinopathy detection. On low-end MCUs, the model experienced a substantial latency of 192,877 ms, making it impractical for time-sensitive tasks. However, on high-end MCUs, latency improved dramatically to 3,948 ms, making it a more feasible option for constrained environments. AI accelerators further enhanced performance, reducing latency to 658 ms, demonstrating their potential for optimizing on-device inference. The best results were seen on traditional computing platforms, with latency dropping to 91 ms on CPUs and just 16 ms on GPUs, as shown in Table 7 and Table 8, highlighting how more powerful hardware can significantly improve real-time processing. These findings indicate that while low-end MCUs may struggle with the demands of the custom DNN, higher-end MCUs, AI accelerators, and GPUs provide viable solutions for real-time diabetic retinopathy screening, depending on the deployment environment and resource availability.

**Table 7: Custom DNN MCU Performance Metrics**

| Device | Latency | EON Compiler RAM | EON Compiler ROM | TFLite RAM | TFLite ROM |
|---|---|---|---|---|---|
| Low-end MCU | 192,877 ms | 2.3M | 1.6M | 4.6M | 1.6M |
| High-end MCU | 3,948 ms | 2.3M | 1.6M | 4.6M | 1.6M |
| AI Accelerator | 658 ms | 2.3M | 1.6M | 4.6M | 1.6M |

**Table 8: Custom DNN Microprocessor Performance**

| Device | Latency | Model Size |
|---|---|---|
| CPU | 91 ms | 1.6M |
| GPU or accelerator | 16 ms | 1.6M |

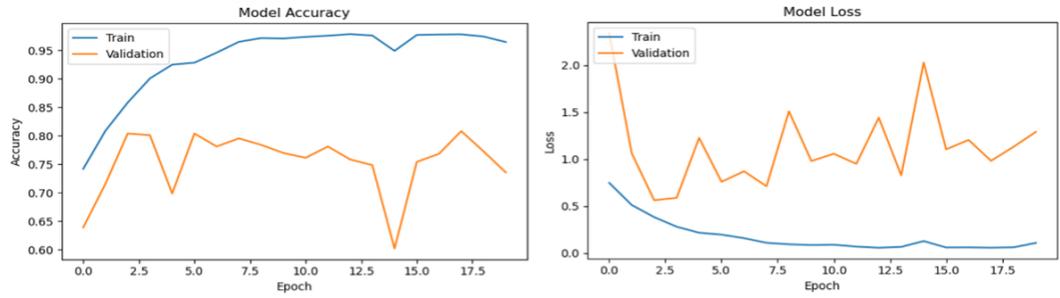

(a)

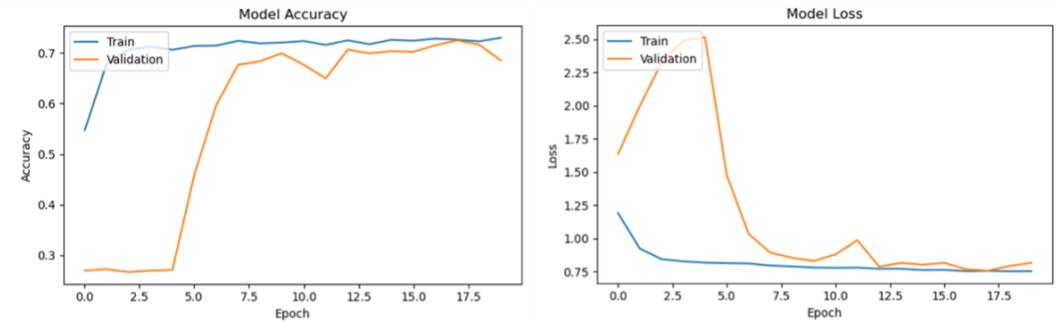

(b)

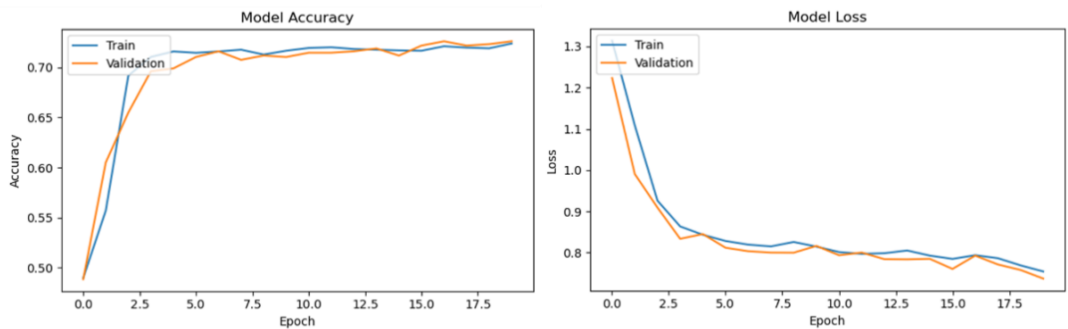

(c)

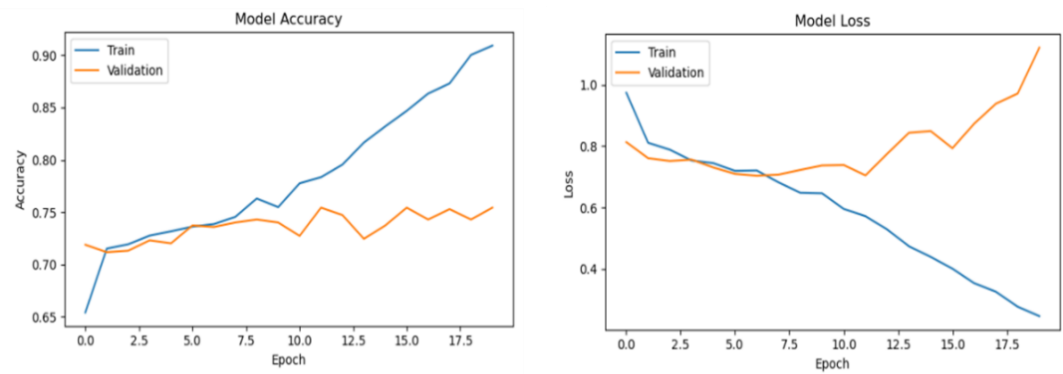

(d)

Figure 11. Graphical recordS of Training accuracy and training loss (a) MobileNet (b) ShuffleNet (c) SqueezeNet (d) Custom DNN.

4.2. **Model Deployment**
In the deployment phase, the trained models were successfully implemented and saved using the Edge Impulse platform in **Figure 12**. As depicted in **Figure 12**, the platform provided a QR code that, when scanned with a mobile device, enabled the smooth transfer and installation of the models onto the target device. This simplified approach ensured that the models could be directly deployed and utilized on the mobile platform, eliminating the need for any further complex setup processes.

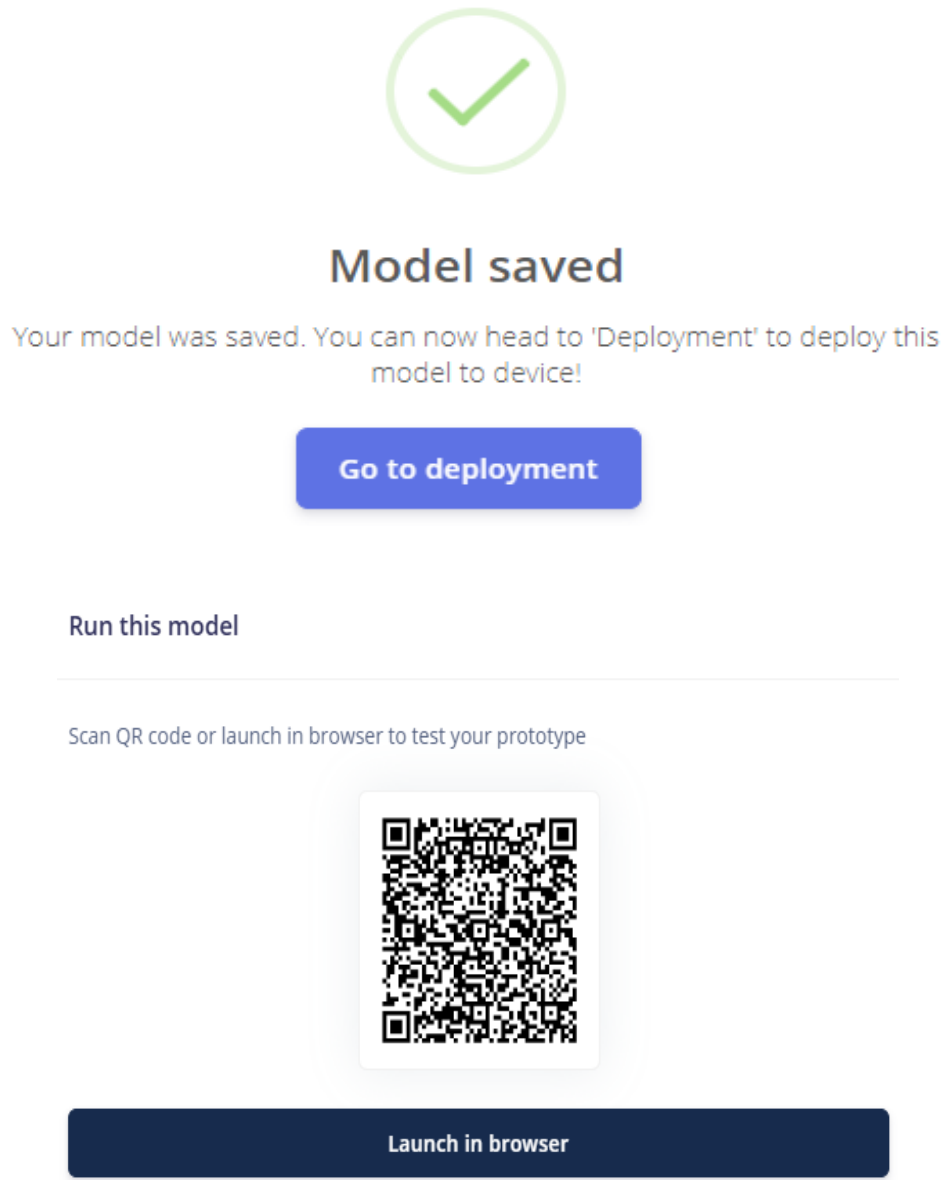

Figure 12. Saved model and model deployment to mobile phone using QR code.

As illustrated in **Figure 13**, the deployed model on a mobile device accurately identifies and classifies the various stages of diabetic retinopathy. This successful prediction underscores the model's ability to differentiate between the distinct health conditions of the retina, demonstrating its effectiveness in detecting and categorizing diabetic retinopathy stages.

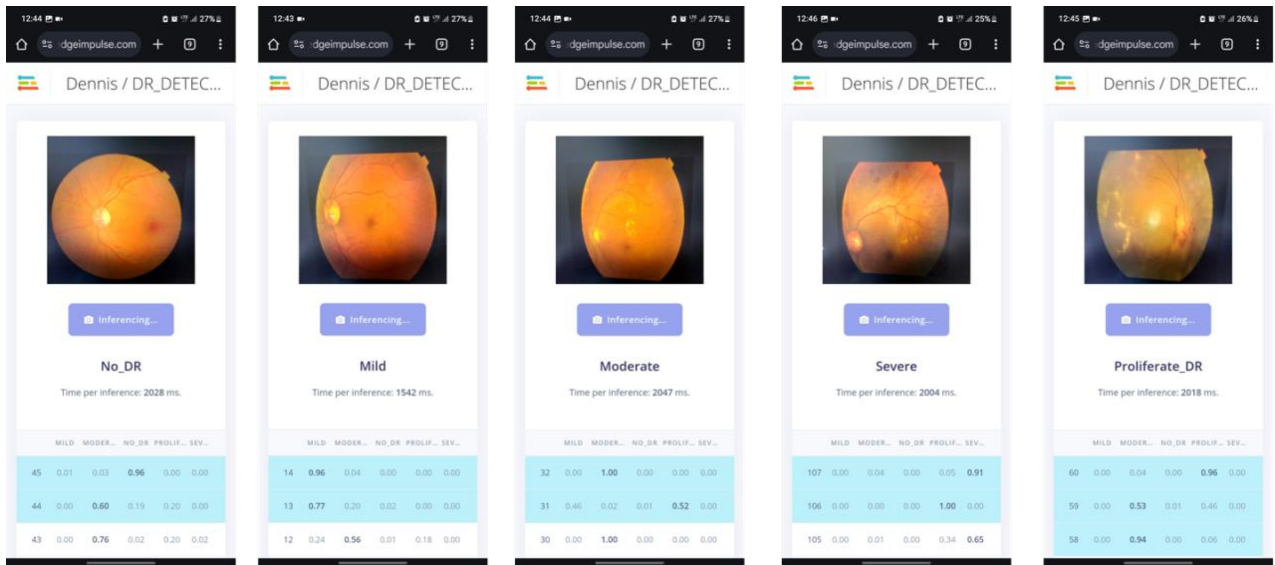
Figure 13. Diabetic Retinopathy detection disease inference on a mobile phone.

## 5. DISCUSSION

The experimental evaluation of deep learning models for diabetic retinopathy detection using the Edge Impulse platform has provided critical insights into their performance on edge devices, with a focus on both accuracy and computational efficiency. Each model exhibited distinct strengths and limitations, especially in the context of resource-constrained environments. MobileNet achieved a high training accuracy (96.45%) and a low training loss (0.1096), reflecting strong performance during the training phase. However, the significant reduction in validation accuracy (73.58%) points to potential overfitting, where the model struggles to generalize on unseen data. Despite its compact model size (3.4MB) and relatively short training time (3,024.75 seconds), the latency observed on low-end MCUs (156,957 ms) suggests that MobileNet is unsuitable for real-time applications in constrained environments. Notably, its performance improves considerably on high-end MCUs (3,214 ms) and AI accelerators (536 ms), highlighting its potential for deployment in settings with more advanced hardware. ShuffleNet demonstrated moderate accuracy, with a training accuracy of 72.96% and a validation accuracy of 68.47%. This reflects a more balanced performance between training and validation, suggesting that it generalizes better than MobileNet. Its small model size (68.5KB) and efficient training time (425.06 seconds) make it a viable option for resource-constrained devices. However, its latency on low-end MCUs (23,169 ms) remains a challenge for real-time deployment. In contrast, the model's performance on high-end MCUs (476 ms) and AI accelerators (80 ms) makes it more suitable for environments where computational resources are less restricted. SqueezeNet offered a similarly balanced performance, with a training accuracy of 72.37% and a validation accuracy of 72.59%. This close alignment suggests the model generalizes well across different datasets. Its small size (176.1KB) and moderate training time (2,427.55 seconds) underscore its efficiency, particularly in environments with limited computational capacity. However, the model's high latency on low-end MCUs (241,592 ms) limits its real-time applicability. Improved results were observed on high-end MCUs (4,946 ms) and AI accelerators (825 ms), positioning SqueezeNet as a potential solution for edge applications in less resource-constrained environments. The custom Deep Neural Network (DNN) exhibited strong training performance, achieving a training accuracy of 90.93%. However, the validation accuracy dropped to 75.43%, indicating some degree of overfitting. The model's moderate size (1.6MB) and training time (2,239.11 seconds) make it a relatively efficient option, especially in comparison to larger architectures. Nonetheless, its latency on low-end MCUs (192,877 ms) was considerable, though performance improved significantly on high-end MCUs (3,948 ms) and AI accelerators (658 ms). This suggests that the custom DNN may be a viable candidate for real-time diabetic retinopathy detection when deployed on more capable edge devices. The findings of this study highlight the trade-offs inherent in deploying deep learning models on edge devices, particularly in balancing accuracy and computational efficiency. Models such as MobileNet and the custom DNN demonstrated high accuracy, but their deployment on low-end MCUs is limited by latency issues. On the other hand, ShuffleNet and SqueezeNet, while less accurate, exhibited better efficiency and lower latency, making them more suitable for resource-constrained environments.

## 5.1 Use Case: Integration of MobileNet into PerceptronCARE for Diabetic Retinopathy Detection in Ghana

In Ghana, diabetic retinopathy (DR) is a leading cause of vision impairment, particularly in rural areas with limited access to specialized care. To tackle this issue, the PerceptronCARE teleophthalmology system has been integrated with the MobileNet deep learning model for early DR detection. **Figure 14** depict the user interface of PerceptronCARE. Healthcare providers in rural clinics can capture retinal images using portable fundus cameras connected to smartphones. These images are uploaded to PerceptronCARE, where the optimized MobileNet model processes them locally, classifying the severity of DR in real-time. The system generates a comprehensive diagnostic report that is accessible to local healthcare professionals for prompt decision-making. If needed, this report can be shared with specialists in urban centers for further review.

The integration of MobileNet offers several benefits:

- Accessibility: Enables early DR screening in remote areas without requiring long-distance travel.
- Real-Time Diagnosis: Provides immediate feedback, reducing delays in intervention.
- Cost-Effectiveness: Operates on low-cost devices, enhancing affordability.
- High Accuracy: Achieves reliable diagnoses, minimizing the risk of misdiagnosis.

Moving forward, PerceptronCARE will focus on optimizing the model with diverse datasets from Ghana and expanding its capabilities to diagnose other conditions, such as cataracts and glaucoma. This initiative exemplifies how AI can address critical healthcare challenges, improving outcomes for diabetic retinopathy patients in Ghana.

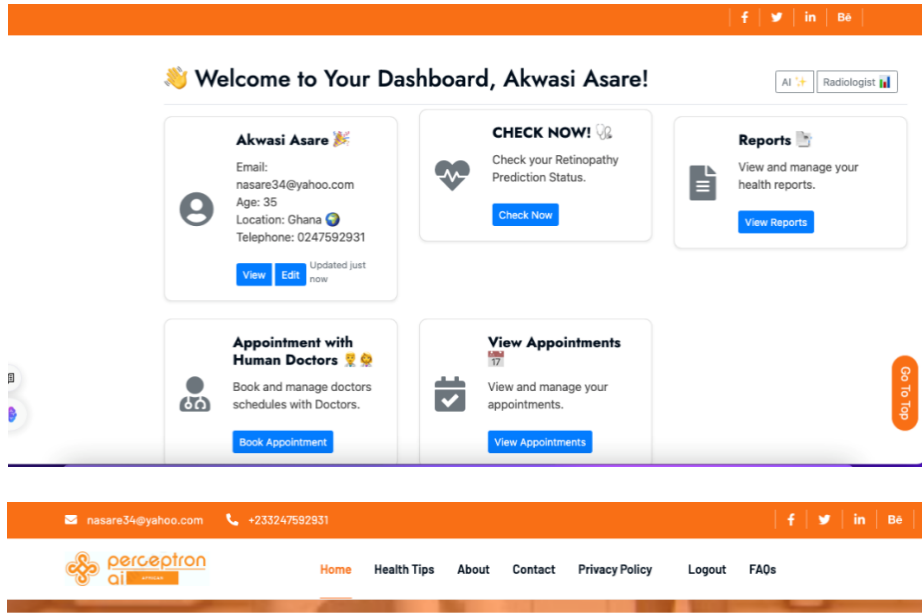
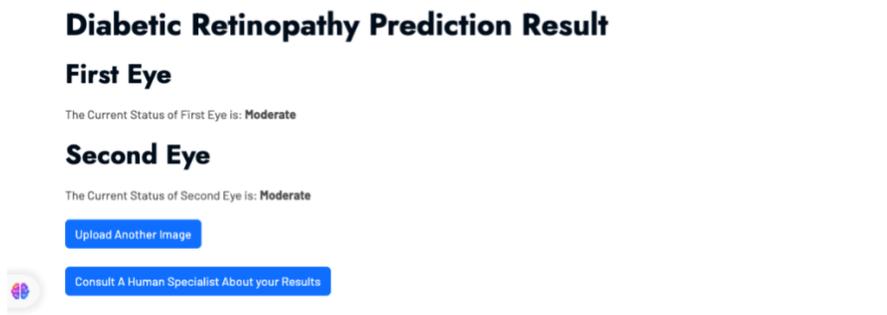

Figure 14. PerceptronCare system.

## 6. CONCLUSION

This study addresses the challenge of diabetic retinopathy detection by utilizing the Edge Impulse platform to deploy various deep learning models, including MobileNet, ShuffleNet, SqueezeNet, and a custom Deep Neural Network (DNN), onto edge devices. This approach enables efficient, real-time detection of different stages of diabetic retinopathy in both clinical and remote healthcare settings. The findings indicate that deploying multiple deep learning models on a single-edge device for diabetic retinopathy detection is a feasible solution, with models such as ShuffleNet and SqueezeNet demonstrating considerable promise due to their balance of performance and computational efficiency. Although MobileNet exhibited high accuracy, further optimization is necessary to address overfitting and reduce latency on resource-constrained devices. The custom DNN, with its effective trade-off between accuracy and resource efficiency, emerges as a strong candidate for edge deployment. Future efforts should prioritize optimizing these models through advanced techniques such as pruning and quantization, aiming to enhance real-time performance and minimize latency. Moreover, expanding and diversifying the training dataset will be essential to improving model robustness and generalization capabilities. Investigating adaptive deployment mechanisms and refining model integration across various edge devices will further ensure their broader applicability in diverse healthcare environments. Additionally, developing user-friendly interfaces will facilitate practical deployment, allowing healthcare professionals to make timely and informed decisions, ultimately improving diabetic retinopathy management in a real-world setting.

## Data Availability Statement

The dataset used in this study is publicly available on Kaggle as part of the Diabetic Retinopathy Detection competition, provided by EyePACS. It can be accessed at [Diabetic Retinopathy Detection - Kaggle](). Any preprocessing steps or modifications applied to the dataset in this study are available upon request from the corresponding author.

## Funding Statement

This research did not receive any financial support from public, private, or non-profit funding agencies.

## Conflict of Interest Disclosure

The authors declare no conflicts of interest related to this study.

## Ethics Approval Statement

This study does not involve human participants or animal subjects; therefore, ethical approval was not required. The dataset used consists of anonymized retinal images provided by EyePACS and made publicly available through Kaggle, ensuring compliance with ethical and legal standards.

## Permission to Reproduce Material from Other Sources

No copyrighted material from third-party sources has been reproduced in this study. All referenced materials are appropriately cited and used in accordance with fair use policies.